\documentclass[12pt,letterpaper]{article}
\usepackage{amsmath}
\usepackage{amsfonts}
\usepackage{bm}
\usepackage[labelfont = bf]{caption}
\usepackage{dsfont}
\usepackage{gensymb}
\usepackage{graphicx,psfrag,epsf}
\usepackage{enumerate}
\usepackage{multirow}
\usepackage{color}
\usepackage{changes}

\usepackage{comment}

\usepackage{natbib}

\usepackage{lineno}

\usepackage{todonotes}

\usepackage{url} 

\newcommand{\blind}{1}

\newcommand{\wh}{\widehat}
\newcommand{\wt}{\widetilde}
\newcommand{\ol}{\overline}

\newcommand{\res}{\text{res}}
\newcommand{\diag}{\text{diag}}

\newcommand{\bu}{\mathbf{u}}

\newcommand{\locset}{\mathcal{ S}}   

\begin{document}

\def\spacingset#1{\renewcommand{\baselinestretch}%
{#1}\small\normalsize} \spacingset{1.35}


\if1\blind
{
  \title{\bf Multivariate postprocessing methods for high-dimensional seasonal weather forecasts}
  \author{Claudio Heinrich, Kristoffer H. Hellton, Alex Lenkoski and \\ Thordis L. Thorarinsdottir\thanks{
    The authors gratefully acknowledge the support of the Volkswagen Foundation through grant nr. 88511 and the Research Council of Norway through grant nr. 270733. We thank Tilmann Gneiting and our project partners in the Seasonal Forecasting Engine project for fruitful discussions.}\hspace{.2cm}\\
    Norwegian Computing Center, Oslo, Norway }
  \maketitle
} \fi

\if0\blind
{
  \bigskip
  \bigskip
  \bigskip
  \begin{center}
    {\LARGE\bf Title}
\end{center}
  \medskip
} \fi

\bigskip
\begin{abstract}
Seasonal weather forecasts are crucial for long-term planning in many practical situations and skillful forecasts may have substantial economic and humanitarian implications. Current seasonal forecasting models require statistical postprocessing of the output to correct systematic biases and unrealistic uncertainty assessments. We propose a multivariate postprocessing approach utilizing covariance tapering, combined with a dimension reduction step based on principal component analysis for efficient computation.  Our proposed technique can correctly and efficiently handle non-stationary, non-isotropic and negatively correlated spatial error patterns, and is applicable on a global scale. Further, a moving average approach to marginal postprocessing is shown to flexibly handle trends in biases caused by global warming, and short training periods. In an application to global sea surface temperature forecasts issued by the Norwegian Climate Prediction Model (NorCPM), our proposed methodology is shown to outperform known reference methods. 
\end{abstract}

\noindent%


\section{Introduction}

Seasonal, or medium-range, weather forecasts on a timescale of one month to a year ahead are highly important in a range of applications. Decisions makers can e.g. greatly benefit from skillful forecasts of increased danger for natural disasters or extreme weather events, such as droughts, hurricanes or extreme snowfall and winds, for efficient mitigation efforts and emergency management. Unlike short-range weather forecasting, medium-range forecasts rely on the prediction of atmospheric modes with a low-frequency variability which can be predicted months ahead. This includes the El Ni\~no Southern Oscillation, monsoon rains and the Northern Atlantic Oscillation \citep{Hoskins2013}. As ocean states change considerably slower than states in the atmosphere, these modes are typically associated with the sea surface temperature in certain regions. Therefore, reliable months-ahead forecasting of sea surface temperature is a crucial first step towards skillful seasonal forecasts of other weather phenomena.  For example, the winter mean surface temperature in large parts of Europe is considered to be negatively correlated with the sea surface temperature in the Nordic seas in the preceding autumn \citep{KolstadAarthun2018,Dobrynin&2018}.

In order to be useful for decision-making, weather forecasts ought to be probabilistic in nature and well calibrated \citep{Gneiting&2007}. Calibration implies that the probability of any event under the forecast distribution matches the actual frequency observed for the event. Current numerical weather prediction (NWP) models are typically deterministic and account for forecast uncertainty by generating an ensemble of forecasts where every ensemble member represents a possible simulation of the future, often generated by creating a small perturbation of the initial state of the prediction.  However, as these models rely on simplifications of the underlying physical system, a (possibly too crude) discretization of space and imperfect initialization, they will be biased. Further, the ensemble spread may not accurately capture the forecast uncertainty. Hence statistical postprocessing is required, where the forecast model is recalibrated based on past performance and observations, see \cite{Vannitsem&2018} for an overview. 

In the postprocessing of medium-range forecasts, obtaining enough training data is a particular challenge. The high-frequency variability patterns need to be filtered out, such that observations have to be averaged over several weeks or months. As the period of reliable sea surface temperature observations starts around 1980, the beginning of the satellite era, the number of available observations for each season is typically below 50. Therefore, medium-range postprocessing techniques must be robust to minimize the risk of overfitting. Additionally, ongoing climate change leads to significant trends in biases and model uncertainty over time \citep{VanSchaeybroeck&2018}. 

Many questions stated by forecast users share the feature that they depend on the forecast distribution at \emph{multiple} locations, so that the forecast must take into account complex dependencies for a skillful prediction of the answers. Examples include predicting the probability of the maximal sea surface temperature in a specific area exceeding 26.5$^\circ$ C, a necessary condition for the development of tropical cyclones \citep{McTaggart&2015}, or predicting the probability of observing sea ice along a shipping route. This requires multivariate postprocessing techniques.

\begin{figure}
\includegraphics[scale = 0.9]{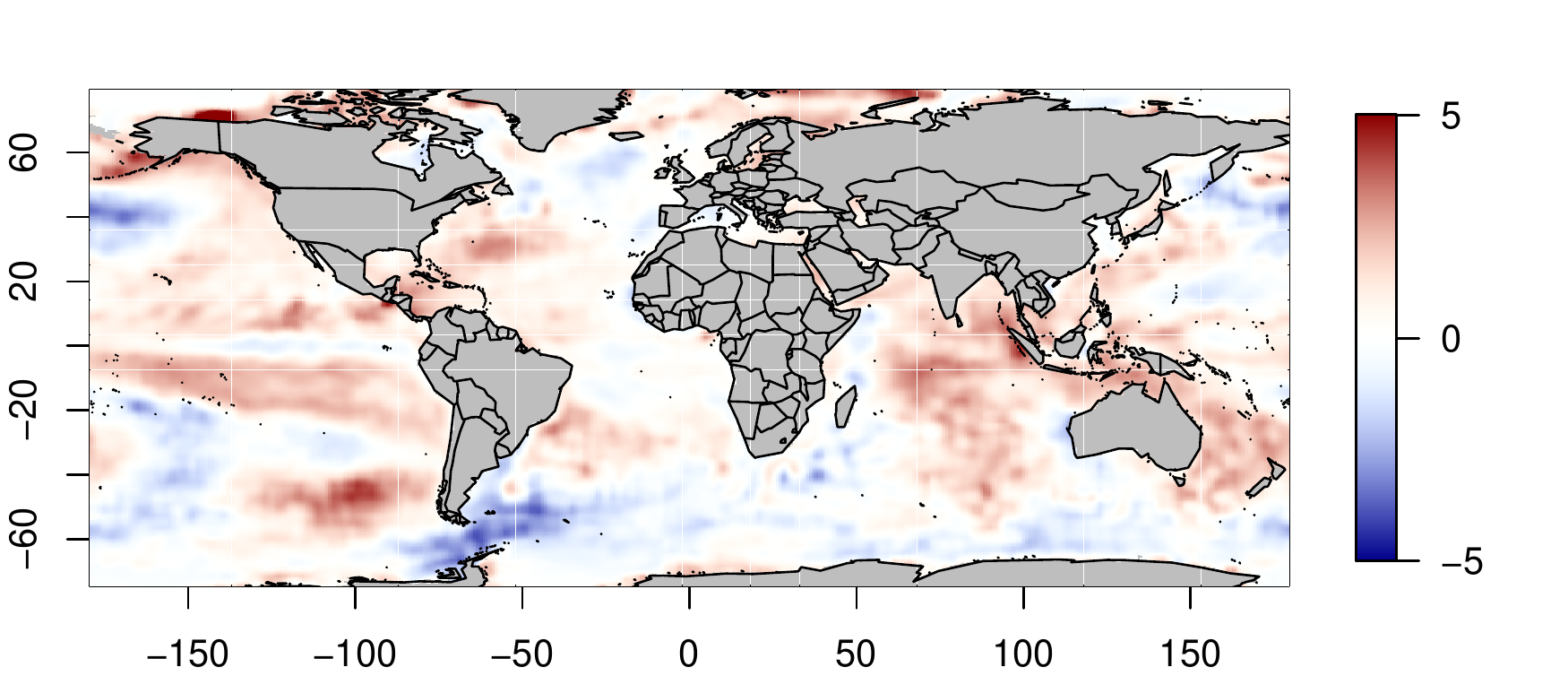}
\caption{Normalized observed sea surface temperature for May 2016 where the normalisation is based on all available observations for May prior to 2016.\label{FigureSST} }
\end{figure}

A common approach to multivariate postprocessing is to fit variograms of a parametric covariance family, such as exponential covariance functions \citep[e.g.][]{Feldmann&2015}. This approach generally assumes stationarity and often nonnegative correlations decaying with distance. Neither of these assumptions are natural when considering global sea surface temperature, as it is likely to depend on ocean currents and the presence or absence of land near, or in between, locations. This is highlighted by Figure \ref{FigureSST} showing normalized observed sea surface temperature for May 2016. Nonstationary effects are visible, such as the strong horizontal correlations in the Pacific ocean westwards of Peru and Columbia due to the El Nino Southern Oscillation. These effects commonly carry forward to the forecast errors in that the model captures the pattern but not the exact magnitude. 

Given the physical complexity underlying the NWP model, forecast residuals of different locations may be negatively correlated.  For example, the sea surface temperature forecast in our data set for July tends to underestimate the temperature in the Baltic sea, when overestimating the temperature in the Barents sea, and vice versa. Most parametric families are not able to model negative correlations, an exception being parametric hole effect models \citep[e.g.][]{ChilesDelfiner1999}. However, these assume that locations at certain distances are always negatively correlated, which is not reasonable in our setup.

We propose a probabilistic multivariate postprocessing approach to tackle these issues and apply it to forecasts of monthly mean sea surface temperature issued by the Norwegian climate prediction model (NorCPM). A moving average approach ensures that the postprocessing will be robust against lack of training data and trends in biases and uncertainties caused by climate change. To achieve spatially consistent forecasts, we explicitly model the spatial dependence structure of the forecast residuals. We utilize regularization of the covariance structure by tapering the covariance matrix, and use further dimension reduction based on principal component analysis to reduce the computational time and reduce the risk of overfitting. Validation on out-of-sample observational data demonstrates that this multivariate postprocessing approach yields spatially consistent and well calibrated forecasts. 

The paper is organized as follows: in Section \ref{Model}, we first develop a univariate postprocessing technique based on moving averages, and extend this to the multivariate setting, incorporating covariance tapering, principal component analysis and a marginal variance correction. Section \ref{Validation} outlines several validation and comparison tools for multivariate forecast distributions. In Section \ref{Results}, we show how the proposed univariate and multivariate methods perform for the NorCPM forecasts and compare their performance to several reference methods. In Section \ref{Case Study}, we consider a case study of a shipping route in the Northern Atlantic, forecasting the probability of ice along the route. The Section \ref{Discussion} gives the concluding remarks and discussion of results. The code for all our methods is available as \verb|R| package at \url{www.github.com/ClaudioHeinrich/pp.sst}.


\section{Modeling sea surface temperature}\label{Model}

\subsection{Data}

We consider monthly mean forecasts of sea surface temperature (SST) issued by the Norwegian climate prediction model (NorCPM), see \cite{Counillon&2014,Counillon&2016}. The forecasts cover the entire globe on a longitude-latitude grid with resolution 1\degree, for a total of 64 800 grid points, where approximately 43 000 are located in the oceans.
NorCPM issues new forecasts every 3 months, at the beginning of January, April, July and October, such that the lead times of the forecasts vary between one and three months. Each forecast consists of nine exchangeable members. For postprocessing, we consider forecasts from 1985 to 2016. The validation period is set to 2001--2016, while the years 1985--2000 are used to train the model. Throughout the validation period, the model estimation is updated for each time point to include the most recent observations. Observations of monthly mean SST over the period are obtained from the Optimum Interpolation Sea Surface Temperature (OISST) dataset of the National Oceanic and Atmospheric Administration \citep{Reynolds&2007}. 
 
\subsection{Univariate postprocessing}\label{Univariate}

A wide range of methods are available for univariate forecast postprocessing,~e.g. ensemble Bayesian model averaging, nonhomogeneous regressions and quantile regression; see \cite{Wilks2018} for a recent overview. In our data set, both bias and prediction uncertainty depend strongly on spatial location and calendar month. Here, we postprocess data from each calendar month separately, ignoring possible interactions between months. In the following, a fixed month is considered with the monthly index suppressed for simplicity. 

For a given year $y \in \{1985,\dots,2016\}$ and location $s\in\locset$, the SST \emph{forecast} is assumed normally distributed, 
\begin{align}\label{model}
X_{y,s}\sim \mathcal N(\mu_{y,s},\sigma_{y,s}^2). 
\end{align} 
The mean and variance are estimated following a (weighted) moving average approach. Specifically, the mean is taken to be the bias-corrected NorCPM ensemble mean, $\wh\mu_{y,s} = \ol{f}_{y,s} - \wh b_{y,s}$, where $\ol f$ denotes the mean of the raw ensemble. To account for trends in the bias over time, for instance caused by climatic changes not accounted for by NorCPM and improved reliability of observations, the bias and predictive variance are estimated by weighted moving averages: 
\begin{align}\label{MA}
\wh b_{y,s} := \sum_{ j<y} w^b_{y-j}(\ol{f}_{j,s} - t_{j,s}), \qquad 
\wh \sigma_{y,s}^2 := \sum_{j<y} w^{\sigma}_{y-j}( t_{j,s}-(\ol{f}_{j,s} - \wh b_{j,s}))^2, 
\end{align}
where $t_{j,s}$ denotes the \emph{observed} temperature at year $j$ and location $s$, and the sequences of weights $w_1,w_2\dots$ are normalized. In Section \ref{Results}, we compare the performance of a simple and an exponentially decaying weighting scheme. 
The weights for the bias are chosen by minimizing the mean squared error (MSE) of the bias-corrected forecast $\ol f - \wh b$. For the variance, they are chosen by minimizing the continuous rank probability score (CRPS). 
The CRPS for predictive distribution $\wh F$ and observation $t$ is defined as
\begin{align}
\textup{CRPS}(\wh F,t) = \mathbb{E}_{\wh F} | X - t| - \frac{1}{2} \mathbb{E}_{\wh F} \mathbb{E}_{\wh F} | X - X' |,\label{crps}
\end{align}
where $X, X' \sim \wh F$ denote two independent random variables. It constitutes a proper scoring rule, cf. \mbox{\citet{GneitingRaftery2007}}, and is often used to assess predictive performance.

A main advantage of this approach is that the weighted averages automatically account for month- and location-specific bias and uncertainty. Therefore, parameters of a weighting scheme may be estimated using all locations and all past months which prevents overfitting and makes optimal use of the available training data.

This comes at the cost that our univariate postprocessing technique is not inherently \emph{coherent}, see \citet{Krzysztofowicz1999, Zhao&2017}, in the sense that it automatically produces forecasts that are no worse than climatology. Coherent models estimate the correlation between forecast and observation and produce skillful forecasts even when this correlation is negative. In our context, if this is done for each location separately it introduces a large risk of overfitting due to the short duration of the training period, whereas estimating the same correlation at all locations makes the estimate too crude. 
The model \eqref{MA} makes the assumption that the NWP forecast indeed contains some signal, which is sometimes not satisfied in seasonal forecasting, see \citet{Schepen&2014}. On the other hand, the model \eqref{MA} only requires to fit the parameters of the weighting scheme. These can be estimated using all locations and months, since for any weighting scheme the bias and variance estimates are month- and location-specific.
As a consequence, the model \eqref{MA} is very robust, which in our context is more valuable than coherence.
Indeed, in Section \ref{univVal} we demonstrate that it outperforms several coherent reference methods.

 Both estimators rely on the ensemble mean only and are independent both of the number of ensemble members and the ensemble spread which is commonly used as predictor of uncertainty in short range weather forecasting \mbox{\citep[e.g.][]{Messner&2017}}.
In seasonal to decadal forecasts the ensemble spread is known to be a less reliable predictor of forecast uncertainty \mbox{\citep{Ho&2013}}. 
This is supported by the findings in Section \ref{Validation}.
Increasing the number of ensemble members usually also benefits the skill of the ensemble mean \mbox{\citep{BuizzaPalmer1998}}
and would decrease the variance of the estimators $\wh b $ and $\wh \sigma$. Therefore, increasing the ensemble size remains desirable in our context, even though it does not directly enter our equations.

In the OISST dataset, the SST is truncated at $-1.79^\circ$C, the assumed freezing temperature of sea water. As this is relevant for relatively few grid points, we apply the same truncation to the predictive distributions after the parameter estimation rather than assuming a truncated normal model in \eqref{model}. In numerical experiments, the truncation error was found to be substantially smaller than the forecast uncertainty. NorCPM, as most climate prediction models, will inherently account for global warming. However, it relies on simplification of the underlying physical processes and is unlikely to fully describe the effects of climate change. Moreover, numerical prediction models, once initialized, tend to drift towards a model attractor which on the seasonal to decadal scale introduces changes in model biases over time. While this may be accounted for with a linear trend term in the bias model \citep{Boer2009}, this was found to reduce the predictive ability here due to overfitting.

The proposed approach is compared against the related and well known non-homogeneous Gaussian regression (NGR) approach \citep{Gneiting&2005}. Here, the mean and the variance are modeled as linear functions of predictor variables, most commonly the ensemble mean and the ensemble spread, i.e. 
\[\wh \mu_{y,s} = a + b \ol{f}_{y,s},\qquad \wh \sigma_{y,s}^2 = c^2 + d^2 S_{y,s}^2,\]
where $S^2$ denotes the sample variance of the forecast ensemble, and $a,b,c,d$ are regression coefficients. The coefficients $a$ and $b$ are fitted by linear regression by minimizing the mean squared error of the forecast, whereas $c$ and $d$ are fitted by minimizing the CRPS over the training period. 
In order to assess sensitivity to month and location, we consider three different versions of NGR for comparison: Grouped by month and location, $NGR_{m,s}$, where the coefficients may depend on both; grouped by location, $NGR_s$; and grouped by month, $NGR_m$. 
In the original reference \citep{Gneiting&2005} $a,b,c$ and $d$ are estimated simultaneously to minimize the CRPS. 
When using the moving average estimates \eqref{MA}, estimating the parameters of the weighting schemes $w^b,w^\sigma$ simultaneously is computationally costly which is why we estimate the weights $w^b$ for the bias correction by MSE-minimization first, and thereafter estimate the variance for the assumed bias correction. In Section \ref{Validation} we assess the skill of the mean model by (out-of-sample) MSE. For this reason we deviate from \citet{Gneiting&2005} and estimate the regression parameters $a,b$ by ordinary least squares.

The main disadvantage of $NGR_{m,s}$ in our context is the vast number of parameters which makes it prone to overfitting. An alternative method is usually referred to as locally adaptive NGR, see for example \cite {ScheuererKoenig2014} . This method avoids fitting $a$ and $b$ for each location while still allowing for location-specific bias of the model which is achieved by considering model and temperature anomalies instead of absolute temperatures. More precisely, defining 
\[\ol{f}^{\text{ano}}_{y,s} := \ol{f}_{y,s} - \frac{1}{\#\{j<y\}}\sum_{j<y}\ol f_{j,s},\qquad t^{\text{ano}}_{y,s} := t_{y,s} - \frac{1}{\#\{j<y\}}\sum_{j<y}t_{j,s},\]
the locally adaptive NGR fits parameters $a,b$ by regressing $t^{\text{ano}}$ on $\ol{f}^{\text{ano}}$, and issues the mean of the predictive distribution as
\[\wh \mu_{y,s} = a + b \ol{f}^{\text{ano}}_{y,s} + t^{\text{clim}}_{y,s},\qquad\text{where}\qquad 
t^{\text{clim}}_{y,s}:= t_{y,s} - t^{\text{ano}}_{y,s}.\]
Unlike the classical NGR this method accounts for the the mean of the predicted and true temperature varying largely between locations, and is used to estimate one pair of regression coefficients $a,b$ for all locations. We denote this reference method by $NGR^{la}_m$, since the parameters $a$ and $b$ are estimated for each month separately and the method is locally adaptive.

\subsection{Multivariate postprocessing}\label{Multivariate}

In order to obtain physically consistent postprocessed forecast fields, the model \eqref{model} must be extended to include spatial correlation.
The main challenge is that the set $\locset$ of considered locations contains around 42 000 points for the entire globe, and is very large compared to the sample size of up to 31 training years. To allow for non-stationary effects and negative correlations, we propose a postprocessing procedure based on regularization of the sample covariance matrix. It relies on a combination of tapering the sample covariance matrix and principal component analysis (PCA). These are classical tools for high-dimensional covariance estimation, but have found little attention in the context of statistical postprocessing of spatial data. As reference methods we compare the proposed technique to a geostationary approach \citep{Feldmann&2015} and ensemble copula coupling \citep{Schefzik&2013}. 


\subsubsection{Postprocessing by regularization of the sample covariance matrix}

The univariate model \eqref{model} is extended by estimating the covariance matrix of the forecast residuals. The residuals are assumed to follow a multivariate normal distribution
\begin{align}\label{resmodel}
\res_y := t_y - \wh\mu_y \sim \mathcal N_S(0,\Sigma_y),\qquad\text{with}\qquad\text{diag($\Sigma_y) = (\wh\sigma_{y,s_i}^2)_{s_i \in \mathcal S}$},
\end{align}
where $\wh \mu_y$ denotes the vector of bias-corrected forecasts $\ol f_y - \wh b_y$, and $t_y$ the vector of observed temperatures.
The covariance matrix $\Sigma_y$ is multi-layered as it captures both the spatial climatological correlation between different locations on the globe and the forecast uncertainty including spatial interactions. 
Given an estimator of the covariance matrix, $\wh \Sigma_y$, a spatial forecast is issued as
\begin{align}\label{spmodel}
X_y\sim \mathcal N_S(\wh \mu_y,\wh \Sigma_y),
\end{align}
generalizing the marginal model in Equation \eqref{model}. In the following, the year and month are assumed fixed and the indices $y$ and $m$ are suppressed.

The standard estimator of the covariance matrix is the sample covariance matrix (SCM):
\[\mathbf{S}(s_i,s_j):= \frac{1}{Y-1} \sum_{y} (t_{y,s_i}-\wh \mu_{y,s_i})(t_{y,s_j}- \wh \mu_{y,s_j}), \]
where the sum runs over all $Y$ previously observed years.
However, in the high-dimensional setting with limited training data, the sample covariance estimator requires regularization. 
We propose a two-step procedure for regularizing $\mathbf{S}$, first applying a distance-dependent tapering, or weighting, of the covariance matrix and secondly utilizing principal component analysis (PCA) to regularize the eigenstructure and reduce dimensionality.

For the first step, the tapering, we consider a positive, monotonically decreasing function $\phi$, defining a spatial correlation function $C_\phi(s_i,s_j) = \phi(\|s_i - s_j\|)$ that only depends on the distance of the locations $ s_i,s_j\in \mathcal S$. The SCM $\mathbf{S}$, is then tapered by $\phi$ by 
\[\mathbf{S}_\phi(s_i,s_j):= \phi(\|s_i-s_j\|)\mathbf{S}(s_i,s_j).\]
The resulting tapered matrix is always positive semi-definite. Tapering covariance matrices by distances is frequently used in atmospheric sciences \citep{GasCohn1999}. \cite{Gneiting2001} argued that the weight function $\phi$ should be twice differentiable with $\phi'(0)=0$ and a minimal second derivative $|\phi''(0)|$, and suggests the function 
\[ \phi_L(t) :=  \phi(t/L),\quad\text{where}\quad \phi(t) := \bigg((1-t)\frac{\sin(2\pi t)}{2\pi t} + \frac{1-\cos(2\pi t)}{2\pi^2 t}\bigg)\mathds 1\{0\leq t\leq 1\}.\]
Here, $\phi$ is supported on $[0,1]$, such that the tapering function $\phi_L$ has a tuning parameter $L$ determining its support. In numerical experiments, the performance of our postprocessing method performed best for $L$ between $1000$km and $4000$km. For the remaining of the paper, $L$ is set to 2500km.

The tapering is beneficial in two ways: Firstly, the SCM does not consider distance between locations, and it will thus have a high risk of spurious correlations given the large number of locations pairs ($\sim 10^{9}$). The spatial correlation is likely to decrease with distance, and the tapering down-weights high correlations between distant locations as these are less credible than those between close locations. Secondly, it removes the rank deficiency of the SCM $\mathbf{S}$ and changes it into a full rank matrix. 
Indeed, the rank of $\mathbf{S}$ is limited by the number of observed years, $Y = 31$, significantly lower than $S\approx$ 42 000. 
The tapered covariance matrix having full rank makes it benefit more from regularization by principal component analysis (PCA).

To further reduce the risk of over-fitting and increase the speed of simulation, PCA is applied to the tapered covariance matrix estimate. PCA can be used to restrict the covariance estimator to a low-dimensional linear subspace with minimal information loss. 
In detail, we consider the eigenvalue decomposition of $\mathbf{S}_\phi$
\[\mathbf{S}_\phi = U \Lambda U^T = \sum_{i=1}^S \lambda_i \bu_i\bu_i^T,\]
with orthogonal eigenvectors $U = [\bu_1,...,\bu_S]$ and eigenvalues $\Lambda = \text{diag}(\lambda_1,...,\lambda_S)$ in decreasing order. 

The ordered eigenvectors, usually referred to as principal components, are orthogonal linear combinations of the locations expressing the highest variance. The underlying assumption is that only the first $d \ll S$ principal components truly represent a signal, whereas the variability of the remaining components represents unstructured noise. Therefore, only the first $d$ eigenvectors are considered
for the covariance estimate: 
\[\wt\Sigma := \sum_{i=1}^d \lambda_i \bu_i\bu_i^T.\]
The truncation of the eigenvalue decomposition will decrease the marginal sample variance at location $s$, the diagonal element ${\mathbf S}_{ss}$, for a given month:
 \begin{align}\label{VarDeficiency}
\wt \Sigma_{ss}\ =\ \sum_{i=1}^d \lambda_i \bu_{is}^2\ <\ \sum_{i=1}^S \lambda_i \bu_{is}^2\ =\ (\mathbf{ S}_\phi)_{ss}\ =\ \mathbf{ S}_{ss}.
\end{align}
Assuming the marginal postprocessing yields calibrated marginal distributions, we want the marginal variances of the multivariate method to equal those estimated by the univariate method. We will therefore compare two alternative approaches for correcting the variance deflation, a multiplicative and an additive correction. In the multiplicative correction, the PCA step is performed on the (tapered) correlation matrix and transformed back to the covariance matrix 
\[\wh\Sigma^{mc}\  :=\ \Xi\,\wt\Sigma\,\Xi,\qquad \text{ with }\qquad\text{$\Xi = \diag\bigg(\frac {\wh\sigma_{y,1}}{(\wt\Sigma_{11})^{1/2}},...,\frac {\wh \sigma_{y,S}}{(\wt\Sigma_{SS})^{1/2}}\bigg)$}, \]
where the marginal variances $\wh \sigma^2_{y,s}$ are estimated as in \eqref{MA}. 

Alternatively, we perform the PCA on the (tapered) covariance matrix and apply an additive correction to the marginal variances,
\[\wh\Sigma^{ac}\  :=\ \wt\Sigma + \diag(\eta_1,...,\eta_S),\qquad\text{where}\qquad \eta_s := \max\{\wh\sigma^2_{y,s} - \wt\Sigma_{ss},0\}.\]
To ensure the positive definiteness of $\wh \Sigma^{ac}$, the difference between the regularized and unregularized marginal variances has to be truncated at zero. The additive correction does not change the off-diagonal elements of $\wt\Sigma$. It, however, only satisfies $\wh\Sigma^{ac}_{ss} = \wh\sigma_{y,s}^2$ for locations $s$ where $\wh\sigma^2_{y,s} \geq \wt\Sigma_{ss}$. As the marginal estimator $\wh\sigma^2_{y,s}$ is not equal to the standard sample variance $(\mathbf{S})_{ss}$, this is not guaranteed for all locations.

The main purpose of applying PCA in this way is to allow for more efficient sampling from the predictive distribution. For instance, when simulating an $S$-dimensional normally distributed $X$ with zero mean and covariance matrix $\wh\Sigma^{mc}$, it is sufficient to simulate a $d$-dimensional vector $Y$ and set
\[X = \Xi\,U^{(d)}\, (\Lambda^{(d)})^{1/2} Y,\]
where $U^{(d)}$ and $\Lambda^{(d)}$ contain the first $d$ eigenvectors and eigenvalues. We found that a dimension reduction with order of magnitude $S/d \approx 100 $ leads to good results. As a consequence, simulating $X$ with (previously computed) covariance matrix $\wh\Sigma^{mc}$ is approximately 100 times faster than simulating from a full rank normal distribution with known Cholesky-decomposition of the covariance matrix, disregarding fixed computation costs. When the additive correction is applied, $X$ is simulated by
\[X = \,U^{(d)}\, (\Lambda^{(d)})^{1/2} Y + \diag(\eta_1,...,\eta_S)Z,\]
where $Y$ is $d$-dimensional standard normal and $Z$ is $S$-dimensional standard normal. This is slower than simulating from the multiplicative corrected covariance estimate, but nevertheless significantly faster than simulating from a general $S$-dimensional normal distribution. 

\subsubsection{Reference methods}

We consider two reference methods commonly used in statistical postprocessing of spatial forecasts, see \cite{Schefzik&2018}. A geostationary approach fits a parametric correlation model, assuming spatial stationarity and isotropy. The parametric correlation function is usually assumed to be in the Whittle-Mat\'ern or the exponential family. \cite{Feldmann&2015} suggest an exponential model with nugget, where the correlation of the forecast error at locations $s_i$ and $s_j$ can be written as
 \[C_{\theta,r}(s_i,s_j)  = (1-\theta)\exp\bigg(-\frac{\|s_i-s_j\|}{r}\bigg) + \theta\delta_{ij}.\]
 The parameters $\theta$ and $r$ are estimated by fitting the variogram of the parametric model to the empirical variogram, for details see \cite{Feldmann&2015}. 

Secondly, we compare to ensemble copula coupling (ECC), see \cite{Schefzik&2013}. The method constructs a postprocessed ensemble of the same size $N$ as the original NWP ensemble. The new ensemble is univariately calibrated and follows the same rank-order structure as the raw NWP ensemble.
This is achieved in two steps: First, a univariately calibrated ensemble $\wt x^{(1)},...,\wt x^{(N)}$ is generated by considering $m$ equally spaced quantiles of the calibrated distribution \eqref{model}, i.e.
\[\wt x^{(i)}_s := F_s^{-1}\bigg(\frac i {N+1}\bigg),\qquad \text{for $i = 1,...,N,$ $s = 1,...,S$,}\]
where $F_s$ denotes the cumulative distribution function of the univariate model \eqref{model} at location $s$. Thereafter, the ensemble indices are permuted at each location to obtain an ensemble with the same rank order structure, or empirical copula, as the raw ensemble forecast. To achieve this, denote by $f^{(i)}_s$ the value of the $i$th ensemble member at location $s$ of the raw forecast and find for each location a permutation $\rho_s$ such that
\[f^{(\rho_s(1))}_s \leq\dots \leq f^{(\rho_s(N))}_s\]
is satisfied. Then, the $i$th member of the multivariate ECC forecast ensemble is 
\[\{\wt x^{(\rho^{-1}_s(i))}_s\}_{s\in\{1,...,S\}}.\]
ECC is computationally efficient and it does not require the specification of a full multivariate distribution. A limitation is that the newly generated ensemble has the same number of members as the NWP ensemble. While extensions to larger ensembles using the rank order structure of historical observations have been proposed \citep[e.g.][]{Schefzik2016}, those do not apply in our setting of limited available historical observations. 

As a third reference method we compare to the Schaake shuffle \citep{Clark&2004}. The Schaake shuffle constructs a postprocessed ensemble that has the same empirical copula as past observations. It works very similar to ECC, except that a post-processed ensemble with the same rank order structure as past observations is constructed rather than with the rank order structure of the raw ensemble forecast. The size of the ensemble created by the Schaake shuffle is therefore not limited by the size of the NWP ensemble but by the number of years in the training data set.


\section{Validation methods}\label{Validation}

We validate predictive performance by assessing the sharpness of the predictive distributions subject to calibration \citep{Gneiting&2007}. Calibration, or reliability, refers to the statistical consistency between the forecast and the observations in the validation period, while sharpness refers to the spread of the predictive distribution. Subject to being calibrated, a sharper forecast is less uncertain and thus more informative.
 
Following \citet{Dawid1984}, probabilistic calibration of marginal forecasts is assessed by the probability integral transform (PIT), i.e. the predictive cumulative distribution function $\wh F$ evaluated at the observation $t$. If $\wh F$ is probabilistically calibrated, the PIT will be uniformly distributed, $\wh F(t) \sim U([0,1])$. To summarize the marginal calibration across grid point locations, we investigate the first two moments of the marginal PIT distribution over all time points in the validation period. A uniform distribution $U([0,1])$ has an expectation of $0.5$ and a standard deviation of $1/(2\sqrt{3}) \simeq 0.29$. It follows that $\mathbb{E}(\wh F(t)) < 0.5$ indicates a positive bias and $\mathbb{E}(\wh F(t)) > 0.5$ indicates a negative bias. If $\textup{SD}(\wh F(t)) < 0.29$, the forecast is overdispersive, and reversely, for $\textup{SD}(\wh F(t)) > 0.29$, it is underdispersive. 

We assess multivariate calibration as proposed by \citet{Thorarinsdottir&2016}. Here, pre-rank functions are employed to map an ensemble of realizations from the multivariate forecast and the observation to real numbers which are subsequently ranked in a standard manner. If the forecast and the observations are statistically indistinguishable, the resulting histogram over the observation ranks is flat, whereas deviations from uniformity indicate miscalibration. \citet{Thorarinsdottir&2016} propose two pre-rank functions which assess the multivariate calibration in slightly different manners. The average pre-rank function finds the average of the marginal univariate ranks while band depth ranking assesses the centrality of the observation within the forecast ensemble as proposed by \citet{Lopez-PintadoRomo2009}. 

Forecast accuracy is typically assessed using proper scoring rules \citep{WinklerMurphy1968, GneitingRaftery2007}. Scoring rules assign a numerical score to each forecast-observation pair, where a lower value indicates better predictive performance. To assess the marginal accuracy, we use the mean squared error (MSE), 
\[
\textup{MSE}(\wh F,t) = \left(\wh\mu - t\right)^{2},
\]
where $\wh \mu$ denotes the mean of $\wh F$. The continuous ranked probability score was defined in \eqref{crps}.
 For an ensemble $\b{x} = \{x_1, \ldots, x_N\}$, the CRPS equals
\[
\textup{CRPS}(\b{x},t) = \frac{1}{N} \sum_{k=1}^N | x_k - t| - \frac{1}{2 N^2} \sum_{k=1}^N \sum_{l = 1}^N |x_k - x_l|.
\]
The MSE is fast to compute and compares different mean models for the predictive distribution. The CRPS provides a more complete picture in that it assesses both calibration and sharpness \citep{GneitingRaftery2007}. 

For a multivariate assessment we utilize the multivariate variogram score (VS) proposed by \citet{ScheuererHamill2015}. For a multivariate distribution function $\wh F$ and an observation vector $\bf{t}$ at $S$ locations, the VS of order $p$ is given by
\[
\textup{VS}(\wh F,{\bf t}) =  \sum_{i=1}^S \sum_{j=1}^S\omega_{ij} \big(|t_i - t_j|^p - \mathbb{E}_{\wh F}|X_i - X_j|^p \big)^2,
\]
where $t_i$ is the observation at the $i$th location and $X_{i}$ the $i$th component of a random vector distributed according to $\wh F$. The (nonnegative) weights $\omega_{ij}$ are set to be constant, such that the correlation structure of all distances is assessed, and we select the order $p=0.5$, as recommended by \citet{ScheuererHamill2015}.

To test significance of score differences, we apply a permutation test relying on resampling \citep{Moeller&2012, good2013permutation}. Two predictive distributions $\wh F_1$ and $\wh F_2$ are compared under a scoring rule $S(F,\cdot)$ using the statistic
\begin{equation}
s:=\frac 1 n \sum_{i=1}^n (S(\wh F_1,y_i)-S(\wh F_2,y_j)) =:\frac 1 n \sum_{i=1}^n S_i.
\label{eq:permuteStatistic}
\end{equation}
The permutation test is then based on resampling copies of $s$ with the labels of $\wh F_1$ and $\wh F_2$ swapped for a random number of summands. Under the null hypothesis, $\wh F_1$ and $\wh F_2$ perform equally well and the permutations would have the same limiting distribution as the statistic, $s$, as $n\to \infty$. By considering the rank of the observed statistic within the permutations, an asymptotic test is obtained. Permutation tests are computationally efficient and, unlike the commonly applied Diebold-Mariano test \citep{DieboldMariano1995}, they do not require the estimation of the asymptotic variance of the score difference $S_i$ which can be involved in the spatial dependence context. 



Let us finally remark that validating forecasts in a high dimensional setting is a challenge in its own rights. For both variogram scores and multivariate rank histograms the role of the forecast dimension has been discussed in the original papers \cite{ScheuererHamill2015,Thorarinsdottir&2016}, and they were found to perform well in dimension $S$ up to 20. When the dimension is much higher than that, and in particular larger than the number of available forecast-observation-couples, new issues arise. For example, the variogram score becomes computationally involved as the number of summands is $S^2$. For multivariate rank histograms, on the other hand, slight misspecifications of the predictive marginal distributions tend to dominate the appearance of the histogram in very high dimension, making it less informative with regard to the multivariate predictive performance.

3njem\section{Results} \label{Results}

\subsection{Training period}

\begin{figure}
\includegraphics[width= \textwidth]{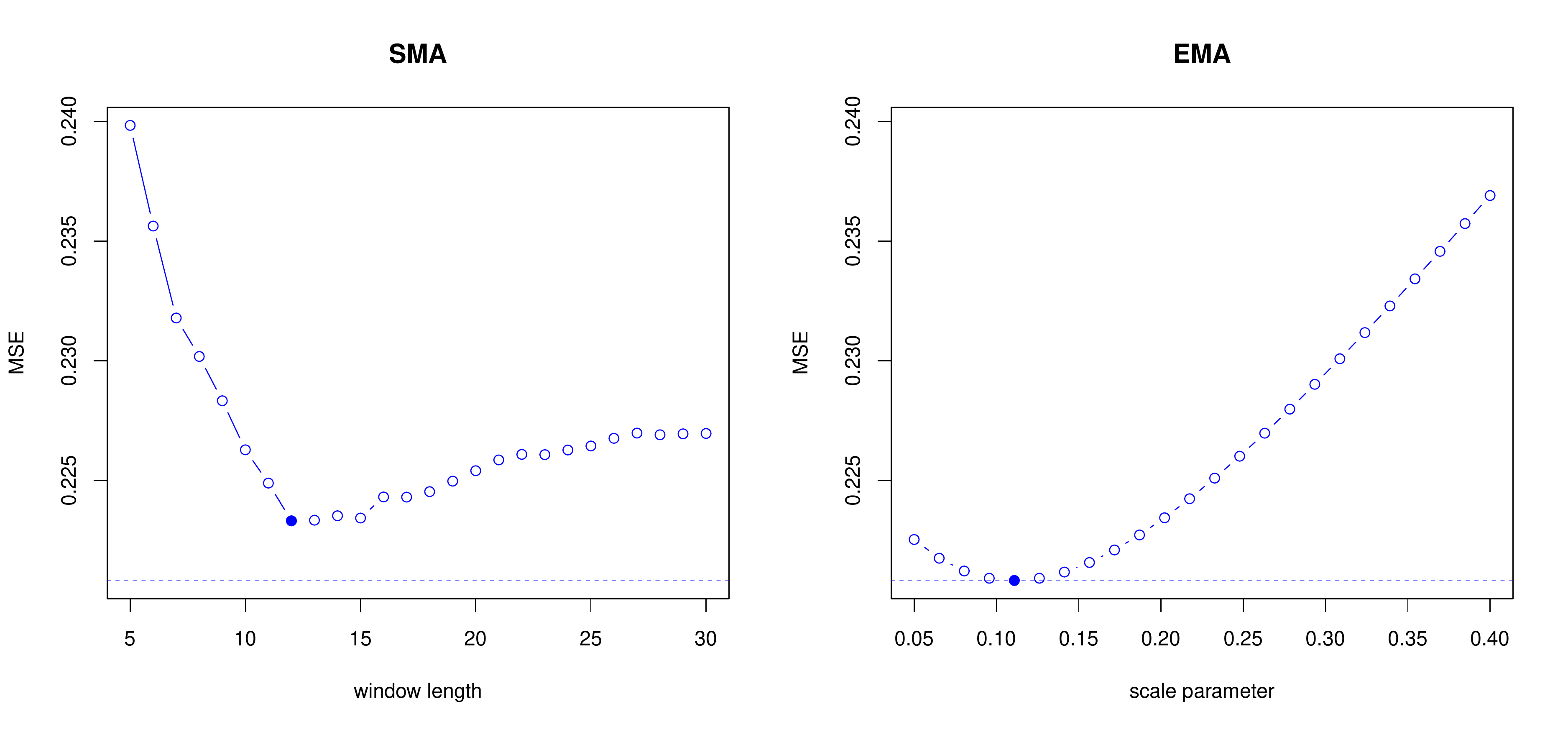}
\caption{Mean squared error (MSE) of bias estimation by simple moving averages (SMA) for varying window length $l$ (left), and for exponential moving averages (EMA) for varying scale parameter $a$ (right), aggregated over all grid points, months and years 2001--2016. The horizontal dashed lines indicate the overall minimum MSE and the filled dots indicate the minimum MSE for each method. The overall minimum is reached for the EMA method with $a = 0.11$.
\label{figMSEMAs} }
\end{figure}

The first step of the analysis is to determine optimal weighting parameters in Equation \eqref{MA}. For simple moving averages and window length $l$, the corresponding weights are $w_k = \mathds 1\{k\leq l\}/l,$ while for exponential moving averages with scale parameter $a$ the weights are $w_k\sim \exp(-ak)$, for the $k$th preceeding year. For the bias-correction, the weighting parameters are chosen for each year in the validation period by minimizing MSE as follows: For a range of weighting parameters, we bias-correct a set of previous forecasts in an out-of-sample fashion and compute the MSE. We then select the weighting parameter for which the MSE is minimized. Figure \ref{figMSEMAs} shows this selection process for 2017. Here, the best performance is obtained for relatively short training periods of $l= 12$ for SMA and $a = 0.11$ for EMA, with a substantial improvement in the predictive performance under EMA. For the variance estimation, the weighting parameters are estimated by minimizing the CRPS. Here, the improvement by using weighted averages is more marginal and unweighted averages lead to close to optimal results. Furthermore, the optimal training period is usually somewhat longer than for the bias-correction. For 2017, the values would be  $l = 28$ for SMA and $a = 0.05$ for EMA. These values are typical for years for which sufficient past training data is available.  

\subsection{Marginal predictive performance}\label{univVal}

For a more formal skill assessment, we compare the aggregated MSE for the SMA and EMA methods against the three NGR reference method. The results are summarized in Table \ref{tableBC}. The $NGR_m$ method performs substantially worse than all others, demonstrating that model biases strongly depend on location. Further, $NGR_{m,s}$ performs significantly better than $NGR_s$, indicating that the bias also varies between seasons. The locally adaptive method $NGR^{la}_{m}$ estimates the same number of parameters as $NGR_m$, which it outperforms clearly, but performs worse than $NGR_{m,s}$.
Both SMA and EMA outperform all other methods, with EMA yielding the overall lowest value as demonstrated in Figure~\ref{figMSEMAs}. 
The second row of the table shows $p$-values of a permutation test assessing significance of the score difference between the model and the best performing model EMA.
The seemingly small score differences are all highly significant, as these values constitute average scores over roughly 8 million score evaluations (192 validation months and 42.000 grid points). 
The $NGR_{m,s}$ model relies on a total of approximately $10^{ 6}$ parameters while the SMA and EMA approaches rely on one parameter each and are thus much more robust towards outliers. 

\begin{table}[h]
 \centering
\begin{tabular}{l | c c c c c c}
Method		& $NGR_m$	& $NGR_s$	& $NGR_{m,s}$	& $NGR^{la}_{m}$ 	& $SMA$ 	& $EMA$\\ \hline\hline
MSE			& 2.028		& 0.417		& 0.227			& 0.331				& 0.223		& {\bf 0.220}\\
$p$-value	& $< 0.1 \%$& $< 0.1 \%$& $< 0.1 \%$	& $< 0.1 \%$		& $< 0.1 \%$& -	
\end{tabular}
\caption{Mean squared error (MSE) over all grid points, months and years in the validation period 2001-2016. $NGR_m$ is linear regression grouped by month, $NGR_s$ by location, and $NGR_{m,s}$ by both, $NGR^{la}_{m}$ is the locally adaptive method described in Section \ref{Univariate}, $SMA$ is bias correction by simple moving averages, and $EMA$ by exponential moving averages. The second row shows $p$-values obtained in a permutation test for the significance of the score difference to the best performing method (EMA).  \label{tableBC}}
 \end{table} 

We continue our analysis using EMA for the bias-correction. Different models for estimating marginal variances are compared in Table~\ref{tableVE} using the CRPS. In particular for the CRPS, miniscule differences can be significant when averaging over many score evaluations.
Permutation tests reveal that even the difference between the mean CRPS for SMA and EMA is highly significant.

In the supplementary material to this article we additionally assess mean and variance estimation for each month separately. This more detailed analysis supports the conclusion that the exponential moving average method performs best overall, in almost all instances significantly better than all other models.

\begin{table}
\centering
\begin{tabular}{l | c c c c c}
Method		& $NGR_m$		& $NGR_s$		& $NGR_{m,s}$		& $SMA$			& $EMA$\\ \hline\hline
CRPS		& 0.2426		& 0.2349		& 0.2311			& 0.2305		& {\bf 0.2304}\\
$p$-value	& $< 0.1 \%$	& $< 0.1 \%$	& $< 0.1 \%$		& $< 0.1 \%$	& -
\end{tabular}
\caption{Continuous ranked probability score (CRPS) for different variance estimation methods with bias-correction by EMA and $p$-values for permutation tests comparing with the best possible model (EMA). Results are aggregated over all grid points, months and years in the validation period 2001-2016 and the best models are indicated in bold.}\label{tableVE}
\end{table}

 \begin{figure}
 \centering
 \includegraphics[scale = 0.8]{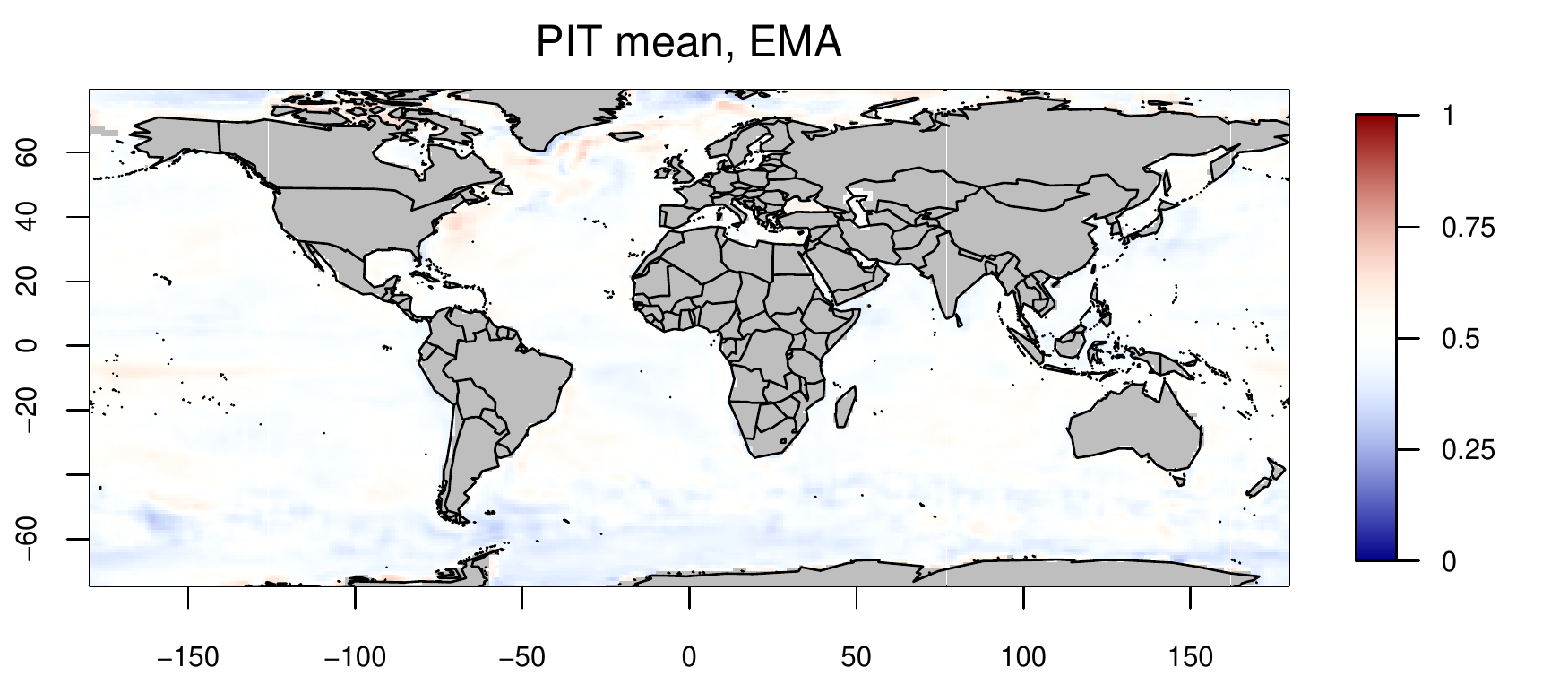}\\[1em]
 \includegraphics[scale = 0.8]{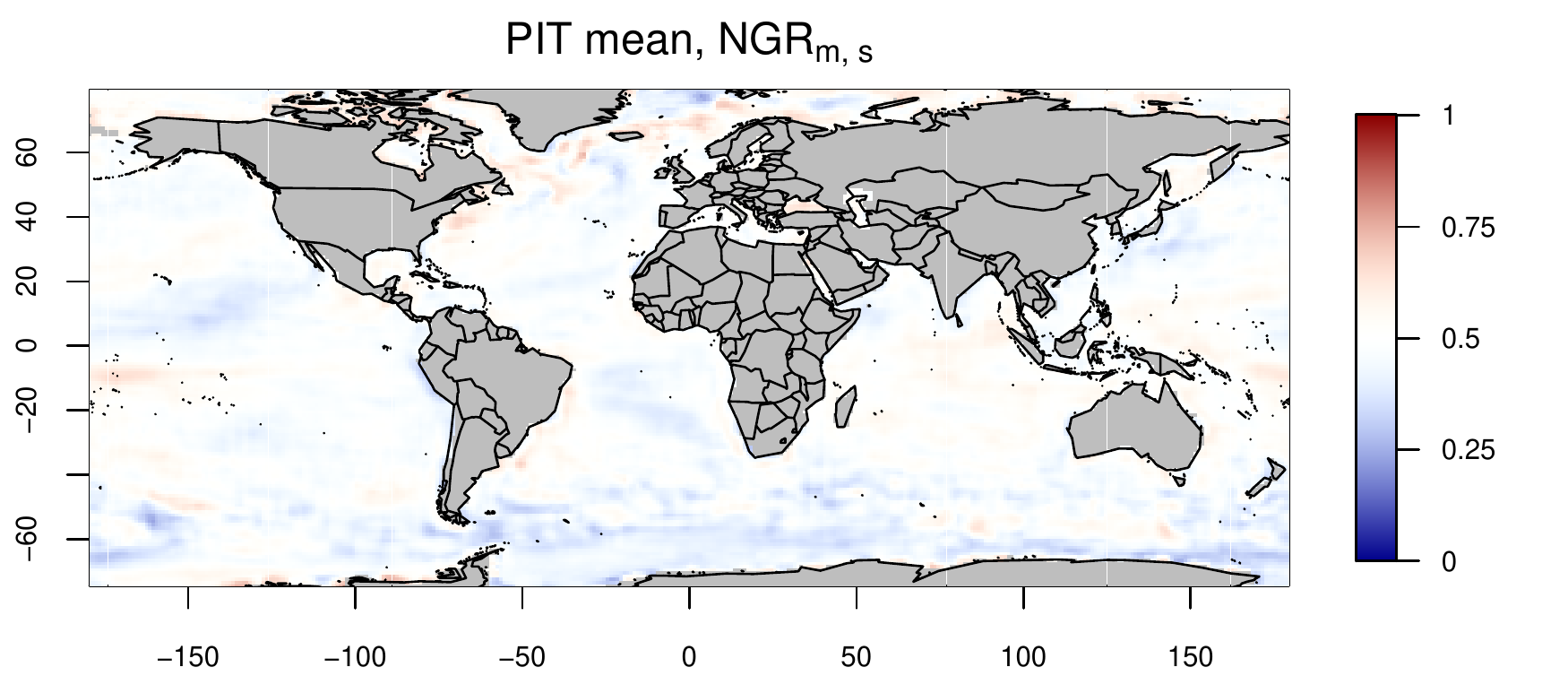}
 \caption{The probability integral transform (PIT) mean values in the validation period 2001--2016 at all locations for EMA (top) and $NGR_{m,s}$ (bottom). White color corresponds to the mean of a uniform random variable, indicating a calibrated forecast. Red shaded areas indicate a negative bias, while blue shaded areas indicate a positive bias. \label{FigurePITs} }
 \end{figure}
 
\begin{figure}
\centering
\includegraphics[scale = 0.8]{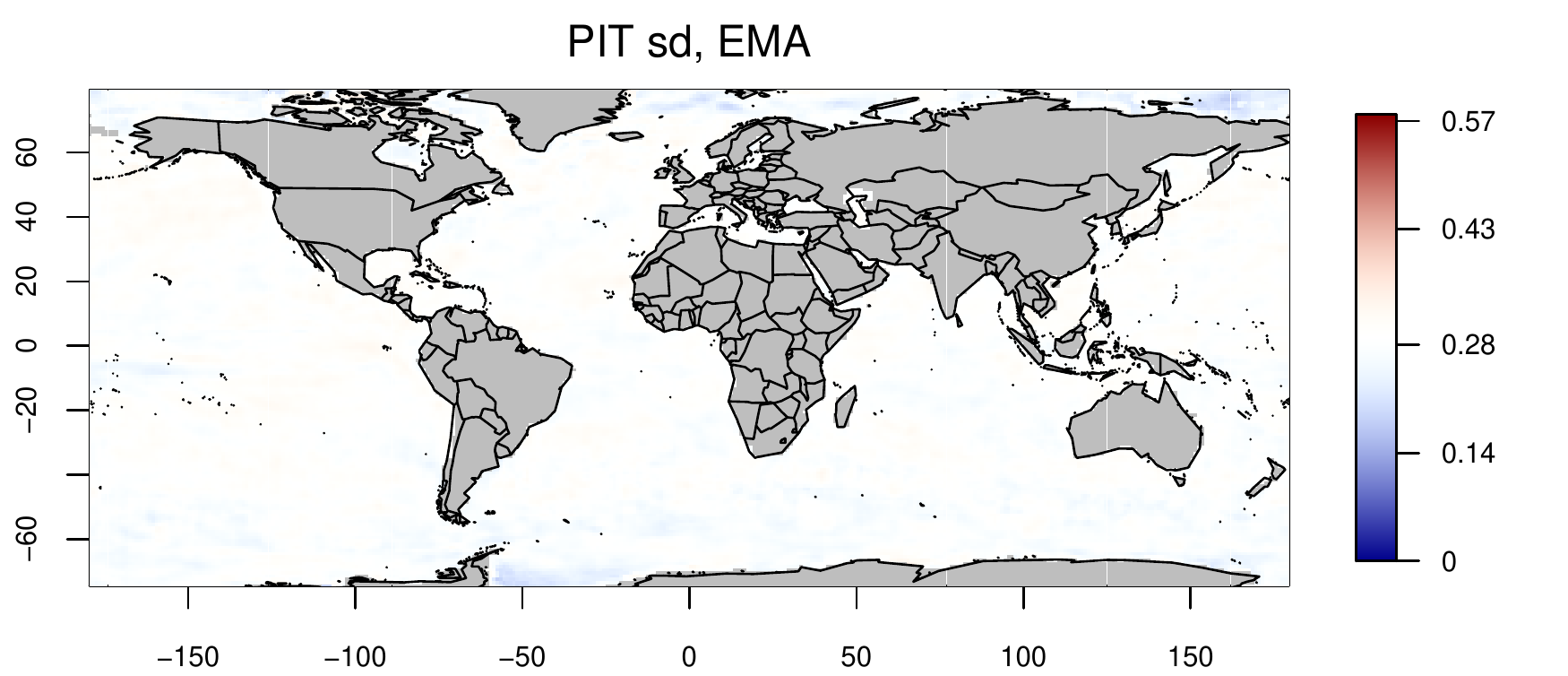}\\[1em]
\caption{The probability integral transform (PIT) standard deviation in the validation period 2001--2016 at all locations for EMA. White color corresponds to the standard deviation of a uniform random variable, indicating a calibrated forecast. The red shaded areas indicate underdispersion and the blue shaded areas indicate overdispersion. \label{fig:PITstanddev} }
\end{figure}

The calibration of EMA and the best NGR method, $NGR_{m,s}$ are assessed in Figure~\ref{FigurePITs}. The postprocessed forecasts are overall well calibrated, indicated by a white color. However, the PIT values reveal small biases in the region governed by the Gulf stream and in the southern oceans below $-50^\circ$ south. The PIT mean values for the EMA method are between 0.26 and 0.72 for all locations, while they range from 0.23 to 0.78 for the $NGR_{m,s}$. The figure indicates that the $NGR_{m,s}$ method exhibits similar biases as the exponential moving average approach, but tends to have larger biases overall. Figure \ref{fig:PITstanddev} further shows the PIT standard deviations across locations for EMA, indicating overall good calibration except in the polar regions where the forecast is somewhat overdispersed. 

\begin{figure}
\center
\includegraphics[width= 0.75\textwidth]{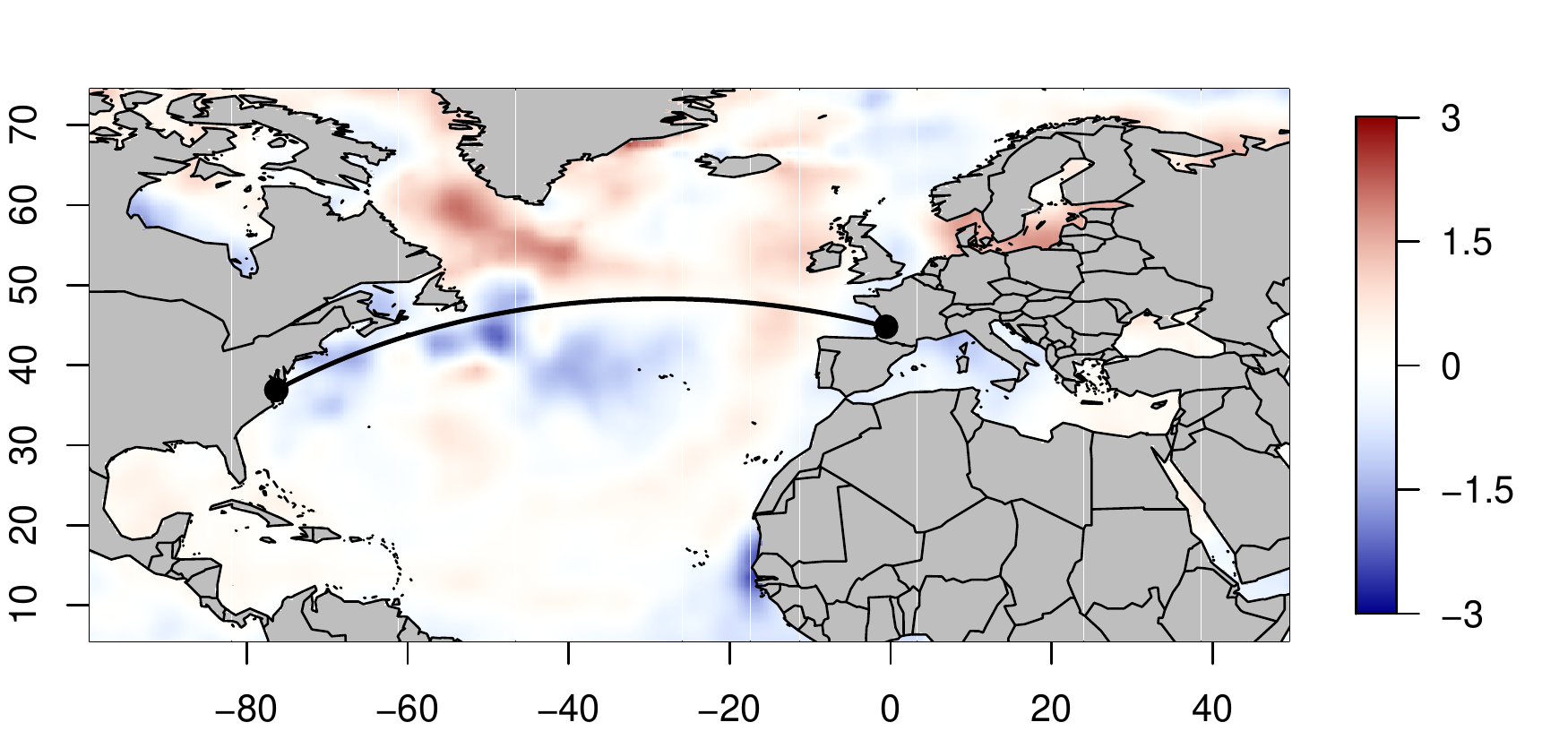}
\caption{Observed forecast residuals in the North-Atlantic for June 2016 with the forecast issued at a lead time of 3 months. The geodesic connecting Norfolk, USA and Bordeaux, France that is considered as shipping route in our case study is shown in black.
 }
\label{figExRes}
\end{figure}

\subsection{Multivariate predictive performance}\label{multivVal}

Here, we compare various multivariate postprocessing approaches where the marginal distributions are generated with EMA. For computational reasons, we restrict our analysis to an area covering the northern half of the Atlantic ocean, cf. Figure \ref{figExRes}. The restricted area covers approximately 5600 grid points. Figure \ref{figExRes} shows the forecast residual, the difference between mean forecast and observations, for June 2016. The aim is for the multivariate correlation structure of the predictive distribution to produce similar spatial patterns. To assess this, we compare the methods in terms of variogram scores. To compute the variogram score, the moments of the predictive distribution, $\mathbb{E}_F[|X_i - X_j|^{1/2}]$ are estimated using 500 simulations from the distribution. For the ECC, the 9 forecast ensemble members are used instead. Table \ref{tabVS} shows the variogram scores averaged over all months in the validation period, with the significance of the score differences assessed by the permutation test. The lowest variogram score is achieved by the regularized covariance matrix with multiplicative correction, $\wh\Sigma^{mc}$, and the score is significantly lower than those for both the geostationary and ECC approach, at a 5\% level. The regularized covariance approach with additive correction, $\wh\Sigma^{ac}$, achieves a similar variogram score as $\wh\Sigma^{mc}$, with a non-significant score difference at the 5\% level. 

\begin{table}
\center
\begin{tabular}{l | c c c c c}
Method:			& $\wh\Sigma^{mc}$		& $\wh\Sigma^{ac}$		& GS			& ECC		& Schaake\\\hline\hline
VS$(F,\bm{y})$	& {\bf 0.03074}			& {\bf 0.03075}			& 0.03410		& 0.03111	& 0.03095\\
$p$-value		& - 					& $40.1 \%$				& $< 0.1 \%$	& $<0.1\%$	& $<0.1\%$
\end{tabular}
\caption{Variogram scores for the area shown in Figure \ref{figExRes} averaged over all months in the validation period 2001--2016, and $p$-values for a permutation test comparing to the best performing model. Numbers in bold highlight models that are not significantly worse than the best performing model, at a level of 5\%.}\label{tabVS}
\end{table}

An empirical assessment of simulated residulas (not shown) suggests that the regularization approach produces the most realistic spatial structure. The residuals generated by the geostationary approach look somewhat too coarse. This is likely caused by an overall poor fit of the parametric variogram to the empirical variogram, resulting in an overestimation of the nugget. The residuals generated by ECC, on the other hand, seem to vary too little on a large scale, which is mainly caused by the low number of only 9 ensemble members.


\section{Case study} \label{Case Study}

In a further assessment of the multivariate predictive distributions, we take a more applied angle and use the model to predict the minimum SST along a shipping route crossing the Atlantic Ocean from Bordeaux, France to Norfolk, USA, see Figure~\ref{figExRes}. The route has a length of 6205 km and we consider all grid cells that are intersected by the geodesic from Bordeaux to Norfolk a part of the route, a total of 93 grid cells. The minimum SST along this route depends jointly on the SST at all locations along the route, requiring spatially coherent forecasts. We consider the same methods of multivariate postprocessing as in the previous section, i.e. the regularization approach, both with multiplicative and additive correction of the marginal variance, as well as the geostationary model and ECC as reference.

\begin{table}
\centering
\begin{tabular}{l | c c c c c}
Method						&$\wh\Sigma^{mc}$			& $\wh\Sigma^{ac}$	& GS			& ECC 		& Schaake			\\ \hline
MSE							& {\bf 0.692}				& {\bf 0.690}		& 0.726			& 0.708		& {\bf 0.687}		\\
$p$-value MSE				& 26.3 $\%$					& 37.1 $\%$			& 2.1 $\%$		& 0.2 $\%$	& -					\\
CRPS						& {\bf 0.445}				& {\bf 0.443} 		& 0.454			& 0.463		&{\bf 0.446}		\\
$p$-value CRPS				& 19.0 $\%$					& -					& 1.2 $\%$		& $<0.1 \%$	& 22.5$\%$			
\end{tabular}
\caption{Scores for minimum SST forecasts along a shipping route from Bordeaux, France to Norfolk, USA for the four different multivariate models, aggregated over all months and years in the validation period 2001--2016. 
 }
\label{figRouteScores}
\end{table}
For each method, we generate multiple simulations from the predictive distribution for each month of the validation period, and compute the minimum temperature along the route. The empirical distribution of simulated minima is then considered the probabilistic forecast of the minimum temperature along the route. For the regularization and geostationary approaches, we simulate 500 forecasts each, whereas for ECC the postprocessed ensemble containing 9 members is used. The accuracy of the forecasts is evaluated with the univariate scores MSE and CRPS, see Table \ref{figRouteScores}. Permutation tests show that the score differences between the Schaake shuffle and the regularization approaches with multiplicative and additive correction of the marginal distribution are not significant at a level of 5\%,
 whereas both the geostationary model and ECC show lower skill. 

\begin{figure}
\includegraphics[width= 0.95\textwidth]{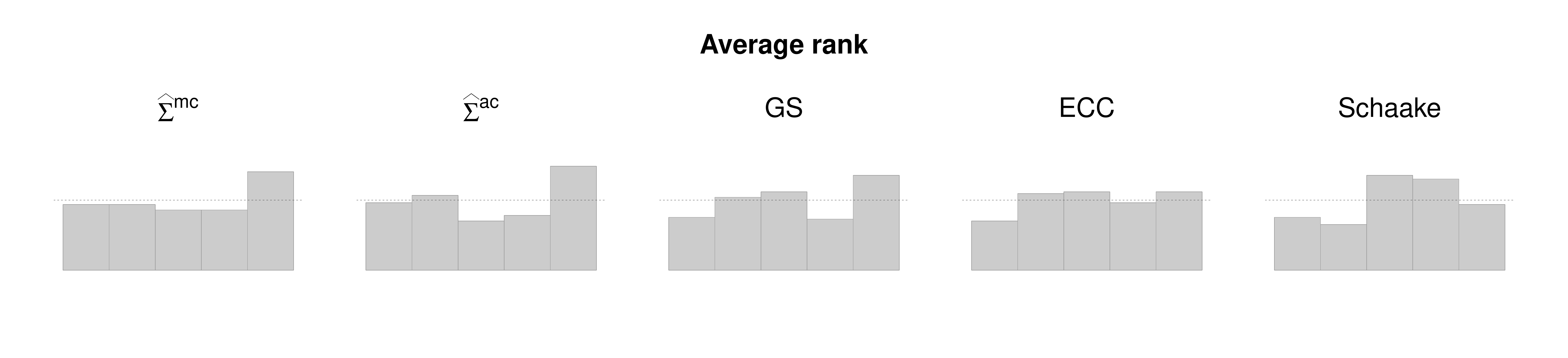}\\[2em]
\includegraphics[width= 0.95\textwidth]{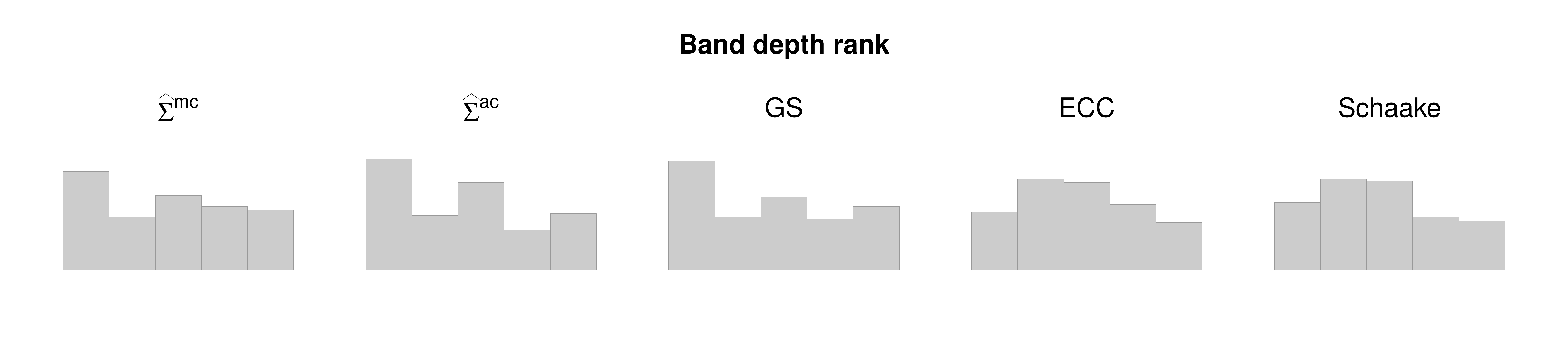}
\caption{Average and band depth rank histograms for the four multivariate methods aggregated over all 192 months in the validation period 2001--2016. The dotted horizontal lines correspond to a perfectly uniform rank histogram.}\label{fig:mvr}
\end{figure}

We further assess the multivariate calibration of the 93-dimensional forecasts using average and band depth ranking, see Figure~\ref{fig:mvr}. Note that the number of available observations in the test set is only 192, making it necessary to restrict the number of bins. The rank histograms of the regularization techniques and the geostationary approach look very similar. In fact, the correlation of the observation ranks is approximately 0.99 for any two of these methods, indicating that deviations from uniformity are mainly attributed to imperfect marginal calibration. All three methods exhibit too many high average ranks, corresponding to an overall underestimation of the temperature along the route, possibly caused by warming trends not accounted for in the numerical model. The same effect is responsible for too many low band depth ranks, signalizing non-centrality of the observation within the ensemble.
The band depth rank histograms for ECC display a slight $\cap$-shape indicating too strong spatial correlations in the empirical copula of the raw NWP ensemble. This is supported by plots of members from the raw ensemble forecast which tend to be visibly smoother than the observation (not shown). 
The band depth rank histogram of the Schaake shuffle looks fairly uniform. If anything, it is slightly $\cap$-shaped which could indicate stronger correlations in the training period than in the validation period, and thus an increasing volatility in the minimum temperature possibly caused by warming.
The average rank histogram for ECC and Schaake shuffle look more uniform than those for the other methods. 
This is presumably caused by the use of equally spaced quantiles which leads to a more evenly spread out predictive ensemble than any ensemble based on simulation. 
However, this comes at the cost of a very limited ensemble size.


\section{Discussion}\label{Discussion}

This paper proposes a fully probabilistic postprocessing approach for multivariate forecasts, the computational costs of which scale well to higher dimensions. The proposed method incorporates a moving average approach combined with regularization of the covariance matrix through tapering. The approach yields a predictive distribution allowing for non-stationary, non-isotropic and negative correlations in the forecasting error. Based on validation data, it performs well with little available training data and is therefore attractive for seasonal and long-range weather predictions.

We have applied the method to seasonal forecasts of sea surface temperature issued by the Norwegian Climate Prediction Model. Performance comparisons indicate that our methodology has higher predictive skill than two reference models, specifically empirical copula coupling and a geostationary method. The Schaake shuffle, our third reference model performed equally good in our case study, but significantly worse when assessed by the variogram score.
The geostationary method assumes the forecast error to have a positive and stationary correlation structure, which in weather forecasting is a highly restrictive assumption, as the underlying physics in numerical prediction models will depend on geographic features not taken into account. Both Schaake shuffle and ECC construct ensembles of fixed size, namely the number of years in the trainings period (16, Schaake shuffle) or the size of the NWP ensemble (9, ECC). The small size of these ensembles restricts the usefulness of these methods for some tasks such as quantile estimation.
This limitation is not shared by our fully probabilistic postprocessing method which can be used to construct ensembles of any size.

There are a variety of regularization methods for sample covariance matrices available in the literature, see \citet{Pourahmadi2013} for an overview.
We believe that our combination of tapering and PCA has several advantages making it appealing in the context of postprocessing seasonal weather forecasts. Firstly, the tapering regularizes by distance, in the sense that the sample covariance of distant gridpoints is down-weighted, but still allows for spatial correlations over long ranges. Given the lack of training data, regularization by distance is more reliable than purely data-driven regularization methods such as e.g. sparse PCA, thresholding or graphical lasso. 
There are some regularization methods that assume sparsity of the precision matrix while accounting for spatial structure of the data, such as for example nearest-neighbour Gaussian fields \citep{Datta&2016}. We expect such methods to lead to similar results as tapering. However, both tapering and PCA have a long tradition in the field of meteorology and are therefore more familiar to practitioners. 
Note that in our model PCA is applied to the already tapered covariance matrix, and its main purpose is not regularization, but reducing the dimension of the predictive distribution. This leads to a massive decrease in the computational costs when sampling from the predictive distribution.

For the covariance tapering we chose the range of our tapering function to be 2500 km. At the equator this corresponds to 22 grid points in each direction. In our experiments, choosing any range between 1000 and 4000 km led to good results and did not affect the conclusions derived from Sections \ref{Results} and \ref{Case Study}. Given the very limited amount of data we refrained from attempting to fit the range parameter. For the PCA, we chose the number $d$ of principal components such that 90\% of the variance where retained. Let us stress that PCA is applied to the already regularized tapered covariance matrix and its main purpose is not regularization, but dimension reduction which allows for faster sampling from the predictive distribution.
When the area with 5.600 gridpoints shown in Figure \ref{figExRes} is considered, 90\% of the variance corresponds to approximately 170 principal components (the exact number varying from month to month). For comparison, if 99\% of the variance should be retained, roughly 600 principal components are required.
It is, in fact, not straightforward to derive an optimal number of principal components based on validation, 
as validating high-dimensional forecasts in itself is involved. An optimization of the variogram score using cross-validation, for example, comes at tremendous computational costs and the score differences are insignificant over a large range of principal components. In other words, the impact of truncating the eigenspace of the tapered covariance matrix by PCA is not picked up by the validation tools we applied. This allows us to choose a relatively small number of principal components in order to lower the computational costs for sampling from the predictive distribution.
The rule of keeping 90\% is therefore a compromise that recovers most of the covariance structure while leading to an increase of sampling speed of factor 20 compared to using the full tapered covariance matrix.

We have investigated an additive and a multiplicative correction of the marginal variances of the multivariate model. The results indicate no significant difference between the skill of these corrections. As the additive correction does not recreate the marginal model \eqref{model} exactly, and is heavier computationally, the approach based on multiplicative correction should be preferred. 

As emphasized by \cite{VanSchaeybroeck&2018}, a main challenge for postprocessing of seasonal weather forecasts is the shortage of available training data. This is supported by the findings in Sections \ref{univVal} and \ref{multivVal}, which demonstrate the risk of overfitting when forecast distributions are estimated separately at each location. We utilized moving averages to estimate the biases and variances. Our approach estimates location-specific biases and variances, but we only need to determine a single, global, weighting parameter. Thus the approach will be more robust against outliers than non-homogeneous Gaussian regression grouped by month and location, the best performing reference model. 
Moreover, our moving average approach will account (to a certain extend) for the trends caused by global warming and the increase in reliability of temperature measurements over the last 30 years.

Future research directions include to consider ensemble information beyond the ensemble mean, for instance the single ensemble members. The good performance of ensemble copula coupling, considering the high number of locations and low number of ensemble members, indicates that the ensemble members do contain valuable information. Ensemble copula coupling relies fully on the empirical copula of the ensemble to capture the multivariate forecast structure, while our approach only considers variability around the ensemble mean.
Combining both sources of information may be a fruitful way to extend postprocessing techniques in the high-dimensional setting. Moreover, we have currently not considered interactions between different months or seasons throughout the year. Early exploratory analyses showed that including information from previous months as predictors did not improved the forecast distribution. It seems, however, reasonable that forecast errors of different months may in fact be correlated and developing detailed models for such interactions may improve the predictive skill. Finally, our model does not account for sea ice in an appropriate way and could be improved by being combined with an external sea ice model.



\end{document}